# A Review of Financial Accounting Fraud Detection based on Data Mining Techniques

Anuj Sharma
Information Systems Area
Indian Institute of Management,
Indore, India

Prabin Kumar Panigrahi
Information Systems Area
Indian Institute of Management,
Indore, India

## ABSTRACT
With an upsurge in financial accounting fraud in the current economic scenario experienced, financial accounting fraud detection (FAFD) has become an emerging topic of great importance for academic, research and industries. The failure of internal auditing system of the organization in identifying the accounting frauds has lead to use of specialized procedures to detect financial accounting fraud, collective known as forensic accounting. Data mining techniques are providing great aid in financial accounting fraud detection, since dealing with the large data volumes and complexities of financial data are big challenges for forensic accounting. This paper presents a comprehensive review of the literature on the application of data mining techniques for the detection of financial accounting fraud and proposes a framework for data mining techniques based accounting fraud detection. The systematic and comprehensive literature review of the data mining techniques applicable to financial accounting fraud detection may provide a foundation to future research in this field. The findings of this review show that data mining techniques like logistic models, neural networks, Bayesian belief network, and decision trees have been applied most extensively to provide primary solutions to the problems inherent in the detection and classification of fraudulent data.

## 1. INTRODUCTION
With an upsurge in financial accounting fraud in the current economic scenario experienced, financial accounting fraud detection (FAFD) have received considerable attention from the investors, academic researchers, media, the financial community and regulators. Due to some high profile financial frauds discovered and reported at large companies like Enron, Lucent, WorldCom and Satyam over the last decade, the requirement of detecting, defining and reporting financial accounting fraud has increased [1].

The Oxford English Dictionary [2] defines fraud as "wrongful or criminal deception intended to result in financial or personal gain." In academic literature fraud is defined as leading to the abuse of a profit organization's system without necessarily leading to direct legal consequences [3]. Although the literature is missing a universally accepted definition of financial fraud, researcher has defined it as "a deliberate act that is contrary to law, rule, or policy with intent to obtain unauthorized financial benefit" [4] and "intentional misstatements or omission of amount by deceiving users of financial statement, especially investors and creditors" [5]. The accounting fraud is executed by making falsified financial accounting statements where the numbers are manipulated by overstating assets, spurious entries related to sales and profit, misappropriation in taxes, or understating liabilities, debts, expenses or losses [1]. The accounting fraud is also defined by accounting professionals as "deliberate and improper manipulation of the recording of data in financial statements in order to achieve an operating profit of the company and appear better than it actually is" [6].

Economically, financial fraud is becoming an increasingly serious problem and effective detecting accounting fraud has always been an important but complex task for accounting professionals [7]. The internal auditing of financial matters in the companies has become an increasingly demanding activity and there are many evidence that 'book cooking' accounting practices are world-wide applied for doing falsified financial frauds [8]. The detection of accounting fraud using traditional internal audit procedures is a difficult or sometimes an impossible task [9]. First, the auditors usually lack the required knowledge concerning the characteristics of accounting fraud. Second, as the fraudulent manipulation of accounting data is so infrequent, most of the auditors lack the experience and expertise needed to detect and prevent frauds. Finally, the other concern people of finance department like Chief Financial Officer (CFO), financial managers and accountants are intentionally trying to deceive the internal or external auditors [10]. While knowing the limitations of an audit, finance and accounting managers have concluded that traditional and standard auditing procedures are insufficient to detect frauds. These limitations of financial auditing suggest the need for additional automatic data analysis procedures and tools for the effective detection of falsified financial statements.

Although the latest revision auditing standards is enlarging auditors' fraud detection responsibility, effective detecting accounting fraud has always been a problem for accounting profession [5]. The failure of internal auditing system of the organization in identifying the accounting frauds has lead to use of specialized procedures to detect financial accounting fraud collective known as forensic accounting. Forensic accounting plays a vital role in detecting these frauds which are difficult to find out in internal auditing by employing accounting, auditing, and investigative skills [11, 12].

Without the fail-safe financial fraud prevention tools and procedures, the accounting fraud has become a business critical problem in current competitive environment. Data Mining based financial fraud detection and fraud control, automates the whole process and helps to reduce the manual work of screening and checking various statements. This area has become one of the key applications of data mining techniques in established industry or government organizations [3].

Data mining is known as gaining insights and identifying interesting patterns from the data stored in large databases in such a way that the patterns and insights are statistically reliable, previously unknown, and actionable [13]. Data mining is also define as "a process that uses statistical, mathematical, artificial intelligence and machine learning techniques to extract and identify useful information and subsequently gaining knowledge from a large database" [14, 15]. The blending point between data mining and detecting accounting fraud is that, data mining as an advanced analytical tool may assist the auditors in decision making and detecting fraud. The data mining techniques have the potential to solve the contradiction between effect and efficiency of fraud detection [5]. Data mining plays an important role in financial accounting fraud detection, as it is often applied to extract and discover the hidden patterns in very large collection of data [7]. An auditor can never become certain about the legitimacy of and intention behind a fraudulent transaction. Concerning this reality, the most optimal and cost effective option is to find out enough evidences of fraud from the available data using specialized mathematical and data processing algorithms [3]. There are many researches that describe the applicability of data mining algorithms in detecting accounting fraud.

As the research about the applicability of data mining techniques for detection of financial accounting fraud is a promising field, many review articles have published in conference proceedings or journal publications. Statistical methods of detecting different types of frauds like credit card fraud, fraudulent money laundering, telecommunications fraud, etc., are reviewed in [16]. Applications of data mining in stock markets and bankruptcy predictions and related fraud detection have been surveyed in [17]. A survey of data mining-based fraud detection research is presented in [3], including credit transaction fraud, telecoms subscription fraud, automobile insurance fraud, terrorist detection, financial crime detection, intrusion and spam detection. Others researcher have reviewed insurance fraud [18] and financial statement fraud [1].

This paper presents a comprehensive review of the research literature on the application of data mining techniques to detect financial accounting fraud. The paper proposes a framework for data mining techniques based accounting fraud detection to help certified public accountants selecting suitable data and data mining technologies for detecting fraud. The systematic and comprehensive literature review of the data mining techniques applicable to financial accounting fraud detection may provide a foundation to future research in this field.

The rest of the paper is organized as follows. Section 2 describes classification of data mining techniques and applications for financial accounting fraud detection. Section 3 provides distribution of the research literature as per the applications and techniques of data mining for the detection of financial accounting fraud. Section 4 describes our framework in more detail. Section 5 outlines future direction of this work.

# 2. CLASSIFICATION OF DATA MINING TECHNIQUES FOR FRAUD DETECTION

In this section, a graphical conceptual framework is proposed for the available literature on the applications of data mining techniques to financial accounting fraud detection. The classification framework, which is shown in Fig. 1, is based on a literature review of existing knowledge on the nature of data mining research [19,20], fraud detection research [1,3,16,17,18].

A classification framework for financial fraud is suggested in [7] based on the financial crime framework of the U.S. Federal Bureau of Investigation [21], which is one of the established frameworks for financial fraud Detection. Fig. 1 consists of two layers, the first comprising the six data mining application classes of classification, clustering, prediction, outlier detection, regression, and visualization [3,16,18,22,23], supported by a set of algorithmic approaches to extract the relevant relationships in the data [14].

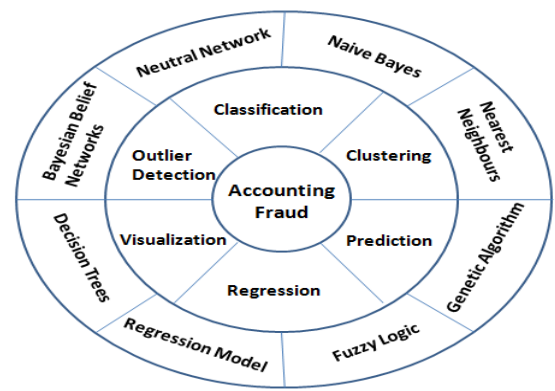

**Fig. 1: The Conceptual Framework for Application of Data Mining to FAFD**

A brief description of the conceptual framework with references is provided and of the six data mining application classes (classification, clustering, outlier detection, prediction, regression and visualization), each component of is discussed in more detail in the following sections.

## 2.1 Classification of Data Mining Applications

Each of the six data mining application classes is supported by a set of algorithmic approaches to extract the relevant relationships in the data. These approaches can handle different classes of problems. The classes are presented below.

**Classification** - Classification builds up (from the training set) and utilizes a model (on the target set) to predict the categorical labels of unknown objects to distinguish between objects of different classes. These categorical labels are predefined, discrete and unordered [24]. The research literature describes that classification or prediction is the process of identifying a set of common features (patterns), and proposing models that describe and distinguish data classes or concepts [17]. Common classification techniques include neural networks, the Naïve Bayes technique, decision trees and support vector machines. Such classification tasks are used in the detection of credit card, healthcare and automobile insurance, and corporate fraud, among other types of fraud, and classification is one of the most common learning models in the application of data mining in fraud detection.

**Clustering** - Clustering is used to partition objects into previously unknown conceptually meaningful groups (i.e. clusters), with the objects in a cluster being similar to one another but very dissimilar to the objects in other clusters. Clustering is also known as data segmentation or partitioning and is regarded as a variant of unsupervised classification [24]. Cluster analysis decomposes or partitions a data set (single or multivariate) into dissimilar groups so that the data points in one group are similar to each other and are as different as possible from the data points in other groups [1]. It is suggested that data objects in each cluster should have high intra-cluster similarity within the same cluster but should have low inter-cluster similarity to those in other clusters [17]. The most common clustering techniques are the K-nearest neighbour, the Naïve Bayes technique and self-organizing maps.

**Prediction** - Prediction estimates numeric and ordered future values based on the patterns of a data set [19]. It is noted that, for prediction, the attribute, for which the value being predicted is continuous-valued (ordered) rather than categorical (discrete-valued and unordered). This attribute is referred as the predicted attribute [24]. Neural networks and logistic model prediction are the most commonly used prediction techniques.

**Outlier Detection** - Outlier detection is employed to measure the "distance" between data objects to detect those objects that are grossly different from or inconsistent with the remaining data set [24]. Data that appear to have different characteristics than the rest of the population are called outliers [26]. The problem of outlier/anomaly detection is one of the most fundamental issues in data mining. A commonly used technique in outlier detection is the discounting learning algorithm [27].

**Regression** - Regression is a statistical methodology used to reveal the relationship between one or more independent variables and a dependent variable (that is continuous-valued) [24]. Many empirical studies have used logistic regression as a benchmark [28]. The regression technique is typically undertaken using such mathematical methods as logistic regression and linear regression, and it is used in the detection of credit card, crop and automobile insurance, and corporate fraud.

**Visualization** - Visualization refers to the easily understandable presentation of data and to methodology that converts complicated data characteristics into clear patterns to allow users to view the complex patterns or relationships uncovered in the data mining process [14]. The researchers have exploited the pattern detection capabilities of the human visual system by building a suite of tools and applications that flexibly encode data using colour, position, size and other visual characteristics. Visualization is best used to deliver complex patterns through the clear presentation of data or functions [29].

## 2.2 Classification of Data Mining Techniques for Financial Accounting Fraud Detection

To determine the main algorithms used for financial accounting fraud detection, we present a Review of data mining techniques identified in literature applied to the detection of financial fraud. The most frequently used techniques are logistic models, neural networks, the Bayesian belief network, and decision trees, all of which fall into the "classification" category. These four techniques are discussed in more detail in the following paragraphs.

**Regression Models** - The regression based models are mostly used in financial accounting fraud detection. The majority of them are based on logistic regression, stepwise-logistic regression, multi criteria decision making method and exponential generalized beta two (EGB2) [7]. Logistic model is a generalized linear model that is used for binomial regression in which the predictor variables can be either numerical or categorical [30, 31]. It is principally used to solve problems caused by insurance and corporate fraud.

Some of the research has suggested logistic regression based model to predict the presence of financial statement fraud [30, 33]. Statistical method of logistic regression can detect falsified financial statements efficiently [30]. Some researchers have also developed generalized qualitative response model based on Probit and Logit techniques to predict financial statement fraud. That model was based on a dataset collected by an international public accounting company and needs testing for generalization [34]. Cascaded Logit model has also proposed to investigate the relationship between insider trading and possibility of fraud. The study in [35] found that, when the fraud is being executed, insiders, i.e. top executives and managers, reduce their stock holdings through high stock selling activity. The other methods like statistical regression analysis are also useful to test if the existence of an independent audit committee mitigates or reduces the likelihood of fraud. Literature also describes that organizations with audit committees, formed by independent managers, meeting no more than twice per year, are less likely to be sanctioned for fraudulent financial reporting [36].

In 2000, Bell and Carcello [37] proposed a logistic regression model for estimating the likelihood of fraudulent financial reporting for an audit client. The model was conditioned on the presence or absence of several fraud-risk factors. The fraud risk factors identified in the final model included weak internal control system, rapid company growth, inadequate or inconsistent relative profitability, management that just want to achieve earnings projections anyhow while lying to the auditors or is overly evasive, company ownership status (public vs. private), and interaction term between a weak control environment and an aggressive management attitude towards financial reporting.

In 2002, Spathis et al. [38] proposed that statistical techniques like logistic regression may be suitable to develop a model to identify factors related to fraudulent financial statement. Non-parametric regression-based framework was used to run the falsified financial statement detection model. The proposed model was compared with discriminant analysis and logit regression methods for benchmarking.

The regression analysis using Logit model can be used for empirical analysis of financial indexes which can significantly predict financial fraud [39]. Logistic analysis and clustering analysis jointly can be used to establish a detecting model of fraud from four aspects of financial indexes, company governance, financial risk and pressure and related trading. After cluster filtering significant variables, prediction model can be established with methods of Standardization, non-Standardization Bayes and Logistic [40].

The logistic regression based accounting fraud detecting models are common in literature since the model based on logistic regression can reach up to 95.1% of detecting accuracy with significant expectation effect [41].

**Neural Networks** – The neural networks are non-linear statistical data modeling tools that are inspired by the functionality of the human brain using a set of interconnected nodes [31, 32]. Neural networks are widely applied in classification and clustering, and its advantages are as follows. First, it is adaptive; second, it can generate robust models; and third, the classification process can be modified if new training weights are set. Neural networks are chiefly applied to credit card, automobile insurance and corporate fraud.

Literature describes that neural networks can be used as a financial fraud detection tool. The neural network fraud classification model employing endogenous financial data created from the learned behavior pattern can be applied to a test sample [42]. The neural networks can be used to predict the occurrence of corporate fraud at the management level [43].

Researchers have explored the effectiveness of neural networks, decision trees and Bayesian belief networks in detecting fraudulent financial statements (FFS) and to identify factors associated with FFS [8].

The study in [10] revealed that input vector consisted of financial ratios and qualitative variables, was more effective when fraud detection model was developed using neural network. The model was also compared with standard statistical methods like linear and quadratic discriminant analysis, as well as logistic regression methods [10].

The generalized adaptive neural network architectures and the adaptive logic network are well received for fraud detection. The hybrid techniques like fuzzy rule integrated with a neural network (neuro-fuzzy systems) are also proposed. The literature describes that the integrated fuzzy neural network outperformed traditional statistical models and neural networks models reported in prior studies [44].

**Bayesian Belief Network** - The Bayesian belief network (BBN) represents a set of random variables and their conditional independencies using a directed acyclic graph (DAG), in which nodes represent random variables and missing edges encode conditional independencies between the variables [8]. The Bayesian belief network is used in developing models for credit card, automobile insurance, and corporate fraud detection. The research in [8] described that Bayesian belief network model correctly classified 90.3% of the validation sample for fraud detection. Bayesian belief network outperformed neural network and decision tree methods and achieved outstanding classification accuracy [8].

**Decision Trees** – A decision tree (DT) is a tree structured decision support tool, where each node represents a test on an attribute and each branch represents possible consequences. In this way, the predictive model attempts to divide observations into mutually exclusive subgroups and is used for data mining and machine learning tasks [8]. Decision trees are predictive decision support tools that create mapping from observations to possible consequences [24]. These trees can be planted via machine-learning-based algorithms such as the ID3, CART and C4.5. Predictions are represented by leaves and the conjunctions of features by branches. Decision trees are commonly used in credit card, automobile insurance, and corporate fraud. To identify and predict the impact of fraudulent financial statements, classification and regression trees (CART) algorithm is introduced in [56].

**Naïve Bayes** - Naïve Bayes is used as simple probabilistic classifier based on Bayes conditional probability rule. Naïve Bayes follows strong (naive) statistical independence assumptions for the predictor variables. It is an effective classification tool that is easy to interpret and particularly suited when the dimensionality of the inputs is high [24]. In a study [3], Naïve Bayes classifier outperformed the conventional classifier. The efficiency of predicting financial fraud data was higher with no false positives with relatively low false negatives. Naïve Bayes methods are widely used in banking and financial fraud detection and claim fraud detection. To apply Ada Boosted Naïve Bayes scoring to insurance claims fraud, a case study is given to diagnosis claim fraud [45].

**Nearest Neighbour Method** - Nearest neighbour method is a similarity based classification approach. Based on a combination of the classes of the most similar k record(s), every record is classified. Sometimes this method is also known as the k-nearest neighbour technique [24]. K-nearest neighbour method is used in automobile insurance claims fraud detection [46] and for identifying defaults of credit card clients [31].

**Fuzzy logic and Genetic Algorithm** – Genetic algorithms are used in classifier systems to represent and modeling the auditor decision behavior in a fraud setting [47]. Genetic algorithm along with binary support vector system (BSVS) which is based on the support vectors in support vector machines (SVM) are used to solve problems of credit card fraud that had not been well identified [48].

Fuzzy Logic is a mathematical technique that classifies subjective reasoning and assigns data to a particular group, or cluster, based on the degree of possibility the data has of being in that group. The expert fuzzy classification techniques enable one to perform approximate reasoning that can improve performance in three ways. First, performance is improved through efficient numerical representation of vague terms, because the fuzzy technology can numerically show representation of a data item in a particular category. The second way performance is enhanced is through increased range of operation in ill-defined environments, which is the way that fuzzy methodology can show partial membership of data elements in one or more categories that may not be clearly defined in traditional analysis. Finally, performance is increased because the fuzzy technology has decreased sensitivity to "noisy" data, or outliers. A multilevel fuzzy rule-based system is proposed in [49] to rank state financial management. The authors used fuzzy set theory to represent imprecision in evaluated information and judgments.

A fuzzy logic model has been implemented in [50] for fraud detection in an Excel spreadsheet. By using the fuzzy logic model to develop clusters for different statements representing red flags in the detection of fraud, non-financial data was included with financial statement variables for the analysis. The model consist of different financial variables like leverage, profitability, liquidity, cash flow and a variable designed to represent a company's risk of fraud. Fuzzy logic efficiently modeled the variable, which was developed to quantify fraud risk factors. The model predicted frauds with 86.7% accuracy [50]. The same model was adapted in [51] to develop a model for detection of financial statement fraud. The proposed model used a combination of different financial statement data. The data included in the model consisted of easily available non-financial information, and a specialized variable constructed by quantifying the company's financial statements issues using fuzzy logic. The study revealed that most successful predictor variables were financial variables indicating cash flow, cash liquidity, the ratio of sales to assets, and a variable indicating company size. The model predicted frauds with 76.7 % accuracy [51].

Fuzzy logic based expert system has been developed to identify and evaluate whether elements of fraud are involved in insurance claims settlements. The fuzzy logic based expert system was developed for auditors to identify fraud in settled claimed insurance. The system was able to cut costs by detecting fraudulent filings [52]. The other fuzzy logic based fraud detection systems are given in [57, 58].

Genetic programming with fuzzy logic production rules is used to classifying data. The study in [53] has proposed and tested a system to detect frauds on real home insurance claims and credit card transaction data. The study on genetic programming for fraud detection lacks benchmarking with the existing methods and techniques. A genetic algorithm based approach to detect financial statement fraud in presented in [54]. It was found that exceptional anomaly scores are valuable metrics for characterizing corporate financial behavior and that analyzing these scores over time represents an effective way of detecting potentially fraudulent behavior. Another study on fraud detection applied many different classification techniques in a supervised two-class setting and a semi-supervised one-class setting in order to compare the performances of these techniques and settings [55].

**Expert Systems** – Researchers in the field of Expert systems have examined the role of Expert Systems in increasing the detecting ability of auditors and statement users. By using expert system, they could have better detecting abilities to accounting fraud risk under different context and level and enable auditors give much reliable auditing suggestions through rational auditing procedure. The research has confirmed that the use of an expert system enhanced the auditors' performance. With assistance from expert system, the auditors discriminated better, among situations with different levels of management fraud-risk. Expert System aided in decision making regarding appropriate audit actions [59].

The financial accounting fraud detection research is classified as per data mining application and data mining techniques. Some researchers have tried to apply a combination of many data mining techniques like decision trees, neural networks, Bayesian belief network, K-nearest neighbour as in [60]. The main objective is to apply a hybrid decision support system using stacking variant methodology to detect fraudulent financial statements. Some research is targeted to indentify financial frauds using simple logistic regression models in specific country like in New Zealand [61] and China [62].

Some of the recent study in the financial fraud detection make used of two or more data mining applications in a hybrid manner or just attempt to compare their effectiveness [70-72].

## 3. THE RESEARCH UNDER PROPOSED CLASSIFICATION FRAMEWORK

The Table 1, 2, 3 and 4 show the distribution of the research papers as per the order of the year of publication.

The Table 1 presents the research work on neural network for financial accounting fraud detection. The literature review is presented in the order of publication. The Table 2 presents the research work on regression models for financial accounting fraud detection. The Table 3 shows the research work on fuzzy logic and Table 4 presents research work on expert system and genetic algorithm for financial accounting fraud detection.

## 4. DATA MINING BASED FRAMEWORK FOR FRAUD DETECTION

The research related with application of data mining algorithms and techniques for financial accounting fraud detection is a well studied area. The implementation of these techniques follows the same information flow of data mining processes in general. The process starts with feature selection then proceeds with representation, data collection and management, pre-processing, data mining, post-processing, and in the end performance evaluation. This paper has proposed an expanded generic data mining framework. This framework considers specific characteristics of fraud detection techniques for financial accounting fraud (see Fig. 2).

## 5. CONCLUSION AND FUTURE RESEARCH

This paper reviewed the literature describing use of data mining algorithms including statistical test, regression analysis, Neural Network, decision tree, Bayesian network etc for financial accounting fraud detection. Regression Analysis is widely used for fraud detection since it has great explanation ability. Different regression model used by researchers are Logit, Step-wise Logistic, UTADIS and EGB2 etc. Neural Networks are important tool for data mining. The researchers have not made any comparison so far, related with detecting effect and accuracy of Neural Network compared to regression model. The advantages of Neural Network are that there are no strict requests for data and it has a strong generalization and adjustment. After correct allocation and proper training, Neural Network may perform great classification comparing with regression model. But due to special inner hidden structure, it is impossible for researchers to track the formation process of output conclusion. There are other issues also related with Neural Network like no clear explanation on connecting weight, complex accuracy and statistical reliability checking procedure, and lack of explanation.

This paper suggests that using only financial statements data may not be sufficient for detections of fraud. The importance of data mining techniques in the detection of financial fraud has been recognized. The future work may be proposing a comprehensive classification framework or a systematic review of data mining application in financial accounting fraud detection. In this study, we conduct an extensive review of academic articles and provide a comprehensive bibliography and classification framework for the applications of data mining to Financial Accounting Fraud Detection. The intention is to inform both academics and practitioners of the areas in which specific data mining techniques can be applied to financial accounting fraud detection, and to report and compile a systematic review of the burgeoning literature on financial accounting fraud detection. Although our study cannot claim to be exhaustive, we believe that it will prove a useful resource for anyone interested in financial accounting fraud detection research, and will help simulate further interest in the field.

**Table 1: Research on Neural Network for Financial Accounting Fraud Detection**

| S. No. | Author | Data Mining Techniques | Main Objective | Reference |
|---|---|---|---|---|
| 1 | Fanning, Cogger and Srivastava (1995) | Neural Networks | To use neural networks to develop a model for detecting managerial fraud | [9] |
| 2 | Green and Choi (1997) | Neural Networks | To develop a neural network fraud classification model employing endogenous financial data in corporate fraud | [42] |
| 3 | Fanning and Cogger (1998) | Neural Networks | To use neural networks to develop a model for detecting managerial fraud | [10] |
| 4 | Cerullo and Cerullo (1999) | Neural Networks | To use neural networks to predict the occurrence of corporate fraud at the management level | [43] |
| 5 | Koskivaara (2000) | Neural Networks | To investigate the impact of various pre-processing models on the forecast capability of neural network for auditing financial accounts | [65] |
| 6 | Feroz et al. (2000) | Neural Networks | To predict the possible fraudsters and accounting manipulations | [66] |
| 7 | Lin, Hwang and Becker (2003) | Fuzzy Neural Network, Logistic Model | To evaluate the utility of an integrated fuzzy neural network model for corporate fraud detection | [44] |
| 8 | Kotsiantis et al. (2006) | Decision Trees, Neural Networks, Bayesian Belief Network, K-Nearest Neighbour | To apply a hybrid decision support system using stacking variant methodology to detect fraudulent financial statements | [60] |
| 9 | Kirkos et al. (2007) | Neural Networks, Decision Trees, Bayesian Belief Network | To explore the effectiveness of neural networks, decision trees and Bayesian belief networks in detecting fraudulent financial statements (FFS) and to identify factors associated with FFS | [8] |
| 10 | Fen-May Liou (2008) | Neural Networks | To build detection/prediction models for detecting fraudulent financial reporting | [68] |
| 11 | M Krambia-Kapardis et al. (2010) | Neural Networks | To test the use of artificial neural networks as a tool in fraud detection | [69] |
| 12 | Ravisankar et al. (2011) | Neural Network, Support Vector Machines | To identify companies that resort to financial statement fraud | [70] |
| 13 | Perols (2011) | Neural Networks, Support Vector Machines | To compares the performance of popular statistical and machine learning models in detecting financial statement fraud | [71] |
| 14 | Zhou and Kapoor (2011) | Neural Networks, Bayesian Networks | To detect financial statement fraud with exploring a self-adaptive framework (based on a response surface model) with domain knowledge | [72] |

**Table 2: Research on Regression Models for Financial Accounting Fraud Detection**

| S. No. | Author | Data Mining Techniques | Main Objective | Reference |
|---|---|---|---|---|
| 1 | Persons (1995) | Logistic Model | To detect financial reporting frauds | [25] |
| 2 | Beasley (1996) | Logit Regression Analysis | To predict the presence of financial statement fraud | [33] |
| 3 | Hansen et al. (1996) | Probit And Logit Techniques | To use Probit and Logit techniques to predict fraud | [34] |
| 5 | Summers and Sweeney (1998) | Logit Regression Analysis | To investigate the relationship between insider trading and fraud | [35] |
| 4 | Abbot, Park and Parker (2000) | Statistical Regression Analysis | To examine if the existence of an independent audit committee mitigates the likelihood of fraud | [36] |
| 6 | Bell and Carcello (2000) | Logistic Model | To develop a logistic regression model to estimate fraudulent financial reporting for an audit client | [37] |

| 7 | Spathis (2002) | Logistic Regression | To use logistic regression to examine published data and develop a model to detect the factors associated with FFS | [30] |
| 8 | Spathis, Doumpos, and Zopounidis (2002) | Logistic Regression | To develop a model to identify factors associated with fraudulent financial statement | [38] |
| 9 | Owusu-Ansah et al. (2002) | Logistic Regression Models | To explore the Logit regression model to detect corporate fraud in New Zealand | [61] |
| 10 | Xuemin Huang (2006) | Regression Analysis Using Logit Model | To analyze financial indexes which can predict financial fraud | [39] |
| 11 | Haisong Ren (2006) | Logistic Analysis And Clustering Analysis | To establish a detecting model of fraud which can be used for empirical analysis of financial indexes | [40] |
| 12 | Guoxin et al. (2007) | Logistic Regression | To develop accounting fraud detecting model | [41] |
| 13 | Bai, Yen and Yang (2008) | Classification And Regression Trees (CART) | To introduce classification and regression trees to identify and predict the impact of fraudulent financial statements | [56] |
| 14 | Yuan et al. (2008) | Logistic Regression Models | To employ a logistic regression model to test the effects of managerial compensation and market competition on financial fraud among listed companies in China | [62] |
| 15 | Fen-May Liou (2008) | Logistic Regression, Classification Trees | To build detection/prediction models for detecting fraudulent financial reporting | [68] |
| 16 | Ravisankar et al. (2011) | Logistic Regression | To identify companies that resort to financial statement fraud | [70] |
| 17 | Perols (2011) | Logistic Regression, C4.5 | To compares the performance of popular statistical and machine learning models in detecting financial statement fraud | [71] |
| 18 | Zhou and Kapoor (2011) | Regression, Decision Tree | To detect financial statement fraud | [72] |

**Table 3: Research on Fuzzy Logic for Financial Accounting Fraud Detection**

| S. No. | Author | Data Mining Techniques | Main Objective | Reference |
|---|---|---|---|---|
| 1 | Deshmukh, Romine and Siegel (1997) | Fuzzy Logic | To provide a fuzzy sets model to assess the risk of managerial fraud | [58] |
| 2 | Deshmukh and Talluru (1998) | Rule-Based Fuzzy Reasoning System | To build a rule-based fuzzy reasoning system to assess the risk of managerial fraud | [57] |
| 3 | Ammar, Wright and Selden (2000) | Fuzzy Logic | To use fuzzy set theory to represent imprecision in evaluated information and judgments | [49] |
| 4 | Lenard and Alam (2004) | Fuzzy Logic and Expert Reasoning | To develop fuzzy logic model to develop clusters for different statements representing red flags in the detection of fraud | [50] |
| 5 | Pathak, Vidyarthi and Summers (2005) | Fuzzy Logic and Expert System | To identify fraud in settled claims | [52] |
| 6 | Chai, Hoogs and Verschueren (2006) | Fuzzy Logic | To convert binary classification rules learned from a genetic Algorithm to a fuzzy score for financial data fraud rule matching | [63] |
| 7 | Lenard, Watkins and Alam (2007) | Fuzzy Logic | To detect financial statement fraud using fuzzy logic | [51] |

**Table 4: Research on Expert System and Genetic Algorithm for Financial Accounting Fraud Detection**

| S. No. | Author | Data Mining Techniques | Main Objective | Reference |
|---|---|---|---|---|
| 1 | Pacheco et al. (1996) | Hybrid intelligent system with NN and fuzzy expert system | To diagnose financial problems in companies | [67] |
| 2 | Eining, Jones and Loebbecke (1997) | Expert System | To build an expert system applying the Logit statistical model to enhance user engagement and | [59] |

| | | | increase reliance on the aid | |
|---|---|---|---|---|
| 3 | Welch, Reeves and Welch (1998) | Evolutionary algorithms (genetic algorithms) | To use genetic algorithms to aid the decisions of Defense Contractor Audit Agency (DCAA) auditors when they are estimating the likelihood of contracts fraud | [47] |
| 4 | Kiehl, Hoogs and LaComb (2005) | Genetic Algorithm | To automatically detect financial statement fraud | [64] |
| 5 | Hoogs et al. (2007) | Genetic Algorithm | To detect financial statement fraud based on anomaly scores as a metrics for characterizing corporate financial behavior | [54] |
| 6 | Juszczak et al. (2008) | Supervised and Semi-Supervised Classification | To detect financial statement fraud | [55] |

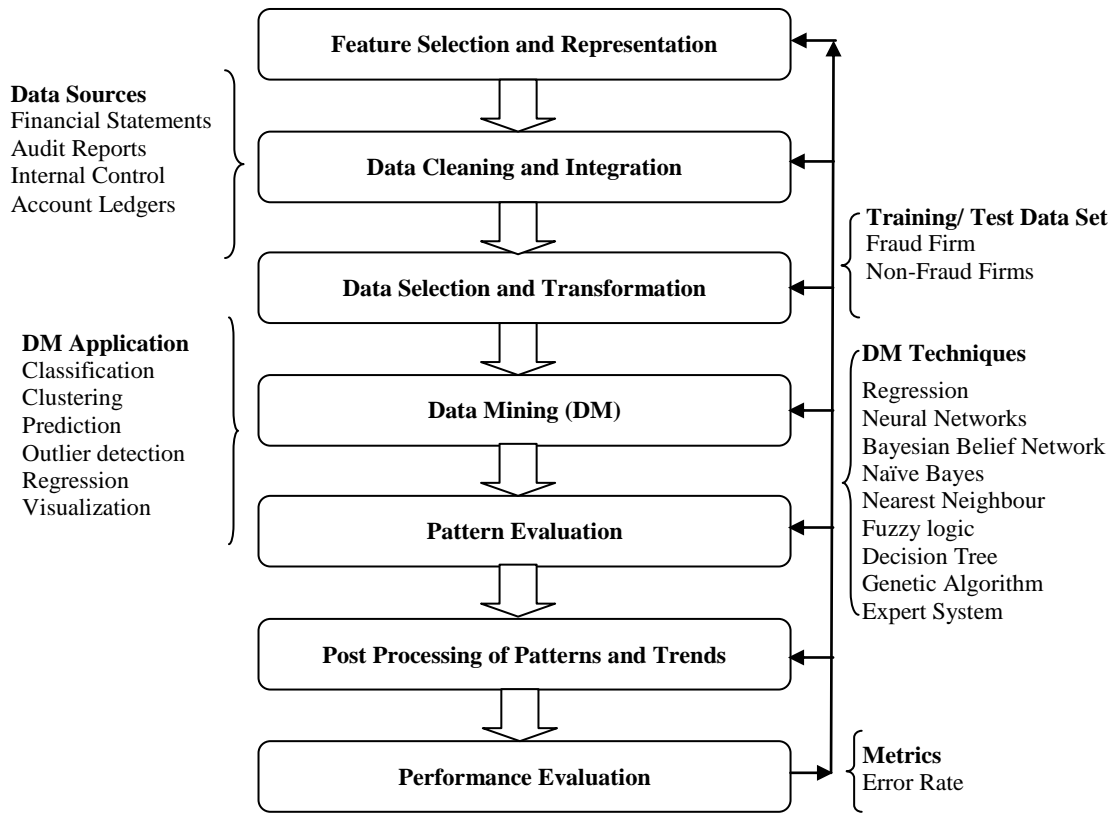

**Fig. 2: The Data Mining Based Framework for Financial Accounting Fraud Detection**

The data mining techniques of outlier detection and visualization have seen only limited use. The lack of research on the application of outlier detection techniques to financial accounting fraud detection may be due to the difficulty of detecting outliers. Indeed, research has point out that outlier detection is a very complex task akin to finding a needle in a haystack. Distinct from other data mining techniques, outlier detection techniques are dedicated to finding rare patterns associated with very few data objects. In the field of financial accounting fraud detection, outlier detection is highly suitable for distinguishing fraudulent data from authentic data, and thus deserves more investigation. Similarly, visualization techniques have a strong ability to recognize and present data anomalies, which could make the identification and quantification of fraud schemes much easier.

The paper suggests that one of the reasons for the limited number of relevant journal articles published for financial accounting fraud detection is the difficulty of obtaining sufficient research data. Fanning and Cogger [9] highlight the challenge of obtaining fraudulent financial statements, and note that this creates enormous obstacles in financial accounting fraud detection research. The most urgent challenge facing financial accounting fraud detection is to bridge the gap between practitioners and researchers. The existing financial accounting fraud detection research concentrates on particular types of data mining techniques or models, but future research should direct its attention toward finding more practical principles and solutions for practitioners to help them to design, develop, and implement data mining and business intelligence systems that can be applied to financial accounting fraud detection.

We predict that increasing amounts of privacy-preserving financial data will be publicly available in the near future due to increased collaboration between practitioners and researchers, and that this should lead to more investigations of data mining techniques that can be applied to privacy-preserving data.

A further problem faced by financial accounting fraud detection is that of cost sensitivity. The cost of misclassification (false positive and false negative errors) differs, with a false negative error (misclassifying a fraudulent activity as a normal activity) usually being more costly than a false positive error (misclassifying a normal activity as a fraudulent activity) [3]. Few studies have explicitly included cost in their financial accounting fraud detection modeling [28], but future research on the application of data mining techniques to financial accounting fraud detection problems should take into account cost sensitivity considerations.

This study has two major limitations. First, our review applied several keywords to search only some online databases for articles published between 1992 and 2011. A future review could be expanded in scope. Second, we considered only articles for financial accounting fraud. Future research could be expanded to include relevant data mining application for detection of frauds in other area like health, communication, insurance, banking etc.

# 6. REFERENCES

[1] Yue, X., Wu, Y., Wang, Y. L., & Chu, C. (2007). A review of data mining-based financial fraud detection research, international conference on wireless communications Sep, Networking and Mobile Computing (2007) 5519–5522.

[2] Oxford Concise English Dictionary, 11th Edition, Oxford University Press, 2009.

[3] Phua, C., Lee, V., Smith, K. & Gayler, R. (2005). A comprehensive survey of data mining-based fraud detection research, Artificial Intelligence Review (2005) 1–14.

[4] Wang, J., Liao, Y., Tsai, T. & Hung, G. (2006). Technology-based financial frauds in Taiwan: issue and approaches, IEEE Conference on: Systems, Man and Cyberspace Oct (2006) 1120–1124.

[5] Wang, S. (2010). A Comprehensive Survey of Data Mining-Based Accounting-Fraud Detection Research. International Conference on Intelligent Computation Technology and Automation, vol. 1, pp.50-53, 2010.

[6] Accounting Fraud Definition and Examples retrieved from http://www.accountingelite.com/accounting-tips/accounting-fraud-definition-and-examples-free-accounting-fraud-article/

[7] Ngai, E.W.T., Hu, Y., Wong, Y. H., Chen, Y., & Sun, X. (2010). The application of data mining techniques in financial fraud detection: A classification framework and an academic review of literature, Decision Support System (2010), doi:10.1016/j.dss.2010.08.006.

[8] Kirkos, E., Spathis, C., & Manolopoulos, Y. (2007). Data mining techniques for the detection of fraudulent financial statements, Expert Systems with Applications 32 (4) (2007) 995–1003.

[9] Fanning, K., Cogger, K., & Srivastava, R. (1995). Detection of management fraud: a neural network approach. International Journal of Intelligent Systems in Accounting, Finance & Management, vol. 4, no. 2, pp. 113– 26, June 1995.

[10] Fanning, K., & Cogger, K. (1998). Neural network detection of management fraud using published financial data. International Journal of Intelligent Systems in Accounting, Finance & Management, vol. 7, no. 1, pp. 21- 24, 1998.

[11] Silverstone, Howard, & Sheetz, M. (2004). Forensic Accounting and Fraud Investigation for Non-Experts. Hoboken, John Wiley & Sons, 2004.

[12] Bologna, Jack & Lindquist, R. J. (1987). Fraud Auditing and Forensic Accounting. New York: John Wiley & Sons, 1987.

[13] Elkan, C. (2001). Magical Thinking in Data Mining: Lessons from COIL Challenge 2000. Proc. of SIGKDD01, 426-431.

[14] Turban, E., Aronson, J.E., Liang, T.P., & Sharda, R. (2007). Decision Support and Business Intelligence Systems, Eighth edition, Pearson Education, 2007.

[15] Frawley, W. J., Piatetsky-Shapiro, G., & Matheus, C.J. (1992). Knowledge discovery in databases: An overview, AI Magazine 13 (3) (1992) 57–70.

[16] Bolton, R. J., & Hand, D.J. (2002). Statistical fraud detection: a review, Statistical Science 17 (3) (2002) 235–255.

[17] Zhang, D., & Zhou, L. (2004). Discovering golden nuggets: data mining in financial application, IEEE Transactions on Systems, Man and Cybernetics 34 (4) (2004) Nov.

[18] Derrig, R. A. (2002). Insurance fraud, The Journal of Risk and Insurance 69 (3) (2002) 271–287.

[19] Ahmed, S.R. (2004). Applications of data mining in retail business, International Conference on Information Technology: Coding and Computing 2 (2) (2004) 455–459.

[20] Mitra, S., Pal, S.K., & Mitra, P. (2002). Data mining in soft computing framework: a survey, IEEE Transactions on Neural Networks 13 (1) (2002) 3–14.

[21] FBI, Federal Bureau of Investigation, Financial Crimes Report to the Public Fiscal Year, Department of Justice, United States, 2007, http://www.fbi.gov/publications/ financial /fcs_ report2007/financial_crime_2007.htm.

[22] Fawcett, T., & Provost, F. (1997). Adaptive fraud detection, Data Mining and Knowledge Discovery 1 (3) (1997) 291–316.

[23] Sánchez, D., Vila, M.A., Cerda, L., & Serrano, J.M. (2009). Association rules applied to credit card fraud detection, Expert Systems with Applications 36 (2) (2009) 3630–3640.

[24] Han, J., & Kamber, M. (2006). Data Mining: Concepts and Techniques, Second edition, Morgan Kaufmann Publishers, 2006, pp. 285–464.

[25] Persons O.S. (1995). Using Financial Statement Data to Identify Factors Associated with Fraudulent Financial Reporting .Journal of Applied Business Research,1995,11(3):38-46.

[26] Agyemang, M., Barker, K., & Alhajj, R. (2006). A comprehensive survey of numeric and symbolic outlier mining techniques, Intelligent Data Analysis 10 (6) (2006) 521–538.

[27] Yamanishi, K., Takeuchi, J., Williams, G., & Milne, P. (2004). On-line unsupervised outlier detection using finite mixtures with discounting learning algorithms, Data Mining and Knowledge Discovery 8 (3) (2004) 275–300.

[28] Viaene, S., Derrig, R.A., Baesens, B., & Dedene, G. (2002). A comparison of state-of-the-art classification techniques for expert automobile insurance claim fraud detection, The Journal of Risk and Insurance 69 (3) (2002) 373–421.

[29] Eick, S.G. & Fyock, D.E. (1996). Visualizing corporate data, AT&T Technical Journal 75 (1) (1996) 74–86.

[30] Spathis, C. T. (2002). Detecting false financial statements using published data: some evidence from Greece, Managerial Auditing Journal 17 (4) (2002) 179–191.

[31] Yeh, I., & Lien, C. (2008). The comparisons of data mining techniques for the predictive accuracy of probability of default of credit card clients, Expert Systems with Applications 36 (2) (2008) 2473–2480.

[32] Ghosh, S., & Reilly, D. L. (1994). Credit card fraud detection with a neural-network, 27th Annual Hawaii International, Conference on System Science 3 (1994) 621–630.

[33] Beasley, M. (1996). An empirical analysis of the relation between board of director composition and financial statement fraud. The Accounting Review, 71(4), 443–466.

[34] Hansen, J. V., McDonald, J. B., Messier, W. F., & Bell, T. B. (1996). A generalized qualitative—response model and the analysis of management fraud. Management Science, 42(7), 1022–1032.

[35] Summers, S. L., & Sweeney, J. T. (1998). Fraudulent misstated financial statements and insider trading: an empirical analysis. The Accounting Review, 73(1), 131–146.

[36] Abbot, J. L., Park, Y., & Parker, S. (2000). The effects of audit committee activity and independence on corporate fraud. Managerial Finance, 26(11), 55–67.

[37] Bell, T., & Carcello, J. (2000). A decision aid for assessing the likelihood of fraudulent financial reporting. Auditing: A Journal of Practice & Theory, 9(1), 169–178.

[38] Spathis, C., Doumpos, M., & Zopounidis, C. (2002). Detecting falsified financial statements: a comparative study using multicriteria analysis and multivariate statistical techniques. The European Accounting Review, 11(3), 509–535.

[39] Huang, X. (2006). Research on Public Company Accounting Fraud and regulation-from perspective of protecting investors [D]. Xiamen: Xiamen University, 2006.

[40] Ren, H. (2006). Investigation Research on Public company financial report fraud. [D] Dalian: Dongbei University of Finance, 2006.

[41] Chen, G., Zhanjia L., & Feng, H. (2007). Empirical Study on detecting financial statements Fraud- based on empirical data of public companies. [J] Auditing Study, 2007.

[42] Green, B. P., & Choi, J. H. (1997). Assessing the risk of management fraud through neural-network technology. Auditing: A Journal of Practice and Theory, 16(1), 14–28.

[43] Cerullo, M. J., Cerullo, V. (1999). Using neural networks to predict financial reporting fraud, Computer Fraud & Security May/June (1999) 14–17.

[44] Lin, J. W., Hwang, M. I., & Becker, J. D. (2003). A Fuzzy Neural Network for Assessing the Risk of Fraudulent Financial Reporting. Managerial Auditing Journal, 2003, 18(8):657-665.

[45] Viaene, S., Derrig, R. A., & Dedene G. (2004). A Case Study of Applying Boosting Naive Bayes to Claim Fraud Diagnosis, IEEE Transactions on Knowledge and Data Engineering, v.16 n.5, p.612-620, May 2004.

[46] Viaene, S., Derrig, R. A., Baesens, & Dedene, G. (2002). A comparison of state-of-the-art classification techniques for expert automobile insurance claim fraud detection, The Journal of Risk and Insurance 69 (3) (2002) 373–421.

[47] Welch, J., Reeves, T. E., & Welch, S. T. (1998). Using a genetic algorithm-based classifier system for modeling auditor decision behavior in a fraud setting, International Journal of Intelligent Systems in Accounting, Finance & Management 7 (3) (1998) 173–186.

[48] Chen, R., Chen, T., & Lin, C. (2006) A new binary support vector system for increasing detection rate of credit card fraud, International Journal of Pattern Recognition and Artificial Intelligence 20 (2) (2006) 227–239.

[49] Ammar, Salwa, Wright, R., & Selden, S. (2000). Ranking State Financial Management: A Multilevel Fuzzy Rule-based System." Decision Sciences. 31: 449-481. 2000.

[50] Lenard, M. J., & Alam, P. (2004). The use of fuzzy logic and expert reasoning for knowledge management and discovery of financial reporting fraud. In H. Nemati and C. Barko (Eds.), Hershey, PA: Idea Group, Inc.

[51] Lenard, M. J., Watkins, A.L., and Alam, P. (2007). Effective use of integrated decision making: An advanced technology model for evaluating fraud in service-based computer and technology firms. The Journal of Emerging Technologies in Accounting 4(1): 123-137.

[52] Pathak, J., Vidyarthi, N., & Summers, S. L. (2005). A fuzzy-based algorithm for auditors to detect elements of fraud in settled insurance claims, Managerial Auditing Journal 20 (6) (2005) 632–644.

[53] Bentley, P. (2000). Evolutionary, my dear Watson: Investigating Committee-based Evolution of Fuzzy Rules for the Detection of Suspicious Insurance Claims. In the Proc. of GECCO 2000.

[54] Hoogs, B., Kiehl, T., Lacomb, C., & Senturk, D. (2007). A genetic algorithm approach to detecting temporal patterns indicative of financial statement fraud, Intelligent Systems in Accounting, Finance and Management, 2007, vol. 15: 41-56.

[55] Juszczak, P., Adams, N.M., Hand, D.J., Whitrow, C., & Weston, D.J. (2008). Off-the-peg and bespoke classifiers for fraud detection", Computational Statistics and Data Analysis, vol. 52 (9): 4521-4532.

[56] Bai, B., Yen, J. & Yang. X. (2008). False financial statements: characteristics of China's listed companies and CART detecting approach, International Journal of Information Technology & Decision Making 7 (2) 339–359.

[57] Deshmukh, A., & Talluru, L. (1998). A rule-based fuzzy reasoning system for assessing the risk of management fraud, International Journal of Intelligent Systems in Accounting, Finance & Management 7 (4) 223–241.

[58] Deshmukh, A., Romine, J., & Siegel, P.H. (1997). Measurement and combination of red flags to assess the risk of management fraud: a fuzzy set approach, Managerial Finance 23 (6) 35–48.

[59] Eining, M. M., Jones, D. R., & Loebbecke, J. K. (1997). Reliance on decision aids: an examination of auditors' assessment of management fraud. Auditing: A Journal of Practice and Theory, 16(2), 1–19.

[60] Kotsiantis, S., Koumanakos, E., Tzelepis, D. & Tampakas, V. (2006). Forecasting fraudulent financial statements using data mining, International Journal of Computational Intelligence 3 (2) 104–110.

[61] Owusu-Ansah, S., Moyes, G.D. , Oyelere, P.B., Hay, P. (2002). An empirical analysis of the likelihood of detecting fraud in New Zealand, Managerial Auditing Journal 17 (4) 192–204.

[62] Yuan, J., Yuan, C., Deng, Y., & Yuan, C. (2008). The effects of manager compensation and market competition on financial fraud in public companies: an empirical study in China, International Journal of Management 25 (2) (2008) 322–335.

[63] Chai, W., Hoogs, B.K., & Verschueren, B.T. (2006). Fuzzy Ranking of Financial Statements for Fraud detection. In proceeding of International Conference on Fuzzy System, (2006), 152–158.

[64] Kiehl, T. R., Hoogs, B. K., & LaComb, C. A. (2005). Evolving Multi-Variate Time-Series Patterns for the Discrimination of Fraudulent Financial Filings. In Proc. of Genetic and Evolutionary Computation Conference, 2005.

[65] Koskivaara, E. (2000). Different pre-processing models for financial accounts when using neural networks for auditing, Proceedings of the 8th European Conference on Information Systems, vol. 1, 2000, pp. 326–3328, Vienna, Austria.

[66] Feroz, E.H., Kwon, T.M., Pastena, V., & Park, K.J. (2000). The efficacy of red flags in predicting the SEC's targets: an artificial neural networks approach, International Journal of Intelligent Systems in Accounting, Finance, and Management 9 (3) 145–157.

[67] Pacheco, R., Martins, A., Barcia, R.M., & Khator, S. (1996). A hybrid intelligent system applied to financial statement analysis, Proceedings of the 5th IEEE conference on Fuzzy Systems, vol. 2, 1996, pp. 1007–10128, New Orleans, LA, USA.

[68] Liou, F. M. (2008). Fraudulent financial reporting detection and business failure prediction models: a comparison. Managerial Auditing Journal Vol. 23 No. 7, pp. 650-662.

[69] Kapardis, M. K., Christodoulou, C. & Agathocleous, M. (2010). Neural networks: the panacea in fraud detection? Managerial Auditing Journal, 25, 659-678.

[70] Ravisankar P., Ravi V., Rao G.R. & Bose I. (2011). Detection of financial statement fraud and feature selection using data mining techniques. Decision Support System, 50, 491-500.

[71] Perols, J. (2011). Financial Statement Fraud Detection: An Analysis of Statistical and Machine Learning Algorithms, Auditing: A Journal of Practice and Theory, Vol. 30(2), pp. 19-50.

[72] Zhou, W. and Kapoor, G. (2011). Detecting evolutionary financial statement fraud. Decision Support Systems, v. 50, n. 3, p. 570-575.